\documentclass[12pt]{iopart}
\usepackage{iopams}

\newcommand{\be}{\begin{equation}}
\newcommand{\ee}{\end{equation}}

\usepackage{iopams,epsf}  
\begin{document}

\title[Exact eigenstates and magnetization jumps in frustrated spin lattices]{
Exact eigenstates and macroscopic
                    magnetization jumps in strongly frustrated
                    spin lattices
}
\author{J. Richter\dag\footnote[3]{To
whom correspondence should be addressed
(johannes.richter@physik.uni-magdeburg.de)}, 
J. Schulenburg\ddag, A. Honecker\S, J. Schnack$\|$ and
H.-J. Schmidt$\|$}  

\address{

\dag Institut f\"ur Theoretische Physik, Universit\"at Magdeburg,
      P.O. Box 4120, D-39016 Magdeburg, Germany

\ddag Universit\"atsrechenzentrum, Universit\"at Magdeburg

\S Institut f\"ur Theoretische Physik, TU Braunschweig

 $\|$ Universit\"at Osnabr\"uck, Fachbereich Physik
}

\begin{abstract}

For a class of frustrated spin lattices including e.g. 
the 1D sawtooth chain, the 2D 
kagom\'e and checkerboard, as well as the 
3D pyrochlore lattices   
we construct exact product eigenstates consisting of several independent,
localized one-magnon states in a ferromagnetic background.
Important geometrical elements of the relevant
lattices are triangles being attached to polygons or lines.
Then the magnons can be trapped on these polygons/lines.
If the concentration of localized magnons is small they can be  distributed
randomly over the lattice. 
Increasing the number of localized magnons their
distribution over the lattice becomes more and more regular and finally
the magnons
condensate in a crystal-like state.

The physical relevance of these eigenstates 
emerges in high magnetic fields where they become groundstates of the
system. As a result a macroscopic magnetization jump appears
in the zero-temperature magnetization 
curve just below the saturation field.
The height of the jump decreases with increasing
spin quantum number and vanishes in the classical limit. Thus it is a
true macroscopic quantum effect.

\end{abstract}

submitted to: {\JPCM}

\section{Introduction}
The search for exact eigenstates of quantum spin systems attracts continuous
attention ever since the Heisenberg and related spin 
models are studied. Of course we
have the fully polarized ferromagnetic state 
as a trivial example.
Furthermore the one- and two-magnon states above the 
fully polarized ferromagnetic state  can be
calculated exactly (see, e.g. \cite{mattis}). A famous 
 example for a very non-trivial eigenstate is Bethe's solution for the
groundstate of the one-dimensional (1D) 
Heisenberg antiferromagnet (HAFM)
\cite{bethe}. 

The investigation of strongly frustrated magnetic systems surprisingly led
to the discovery of several new exact eigenstates.
Whereas in general the treatment of frustrated quantum magnets is even
harder than that of unfrustrated systems in some exceptional cases 
one finds  eigenstates of quite simple nature. The interest in these
eigenstates   comes from the fact that often they become groundstates
either for particular values of frustration or in magnetic fields.
Furthermore the spin correlation functions can be calculated analytically.
Therefore these exact eigenstates play an important role either as 
groundstates of real quantum magnets or at least as reference states
of idealized models for more complex quantum spin systems.
There are two well-known examples for simple eigenstates 
of strongly frustrated
quantum spin systems, namely the Majumdar-Gosh state of the 1D 
$J_1-J_2$ spin 1/2 HAFM \cite{majumdar} and the orthogonal dimer state in the
Shastry-Sutherland  model \cite{shastry}.  Both eigenstates are 
products of dimer singlets and become groundstates only for strong 
frustration. These eigenstates indeed play
a role in realistic materials. While the Majumdar-Ghosh state has some
relevance in quasi-1D spin-Peierls materials like $CuGeO_3$ (see e.g. 
\cite{cugeo}),
the orthogonal dimer state of the Shastry-Sutherland model 
is the magnetic groundstate of the quasi-2d 
SrCu$_2$(BO$_3$)$_2$\cite{srcubo}. 
Other frustrated spin models in one, two or three 
dimensions can be constructed having also dimer singlet product states as
groundstates (see e.g. \cite{pimpinelli,ivanov97,japaner3d}).

Less known is the so-called uniformly distributed
resonating valence bond  state which is the groundstate of the
$J_1-J_2$ chain with ferromagnetic nearest-neighbour (NN) bonds $J_1 <0$ and 
frustrating antiferromagnetic next NN bonds $J_2=-J_1/4$ \cite{hamada}.
  
Another striking feature of the dimer singlet product groundstates is 
the existence of magnetization plateaus at zero magnetization. These plateaus
in quantum spin systems currently attract a lot of attention from the
theoretical as well as experimental side.      
Recently it has been demonstrated for the 1d counterpart of the
Shastry-Sutherland model, the so-called frustrated dimer-plaquette
chain (also known as orthogonal dimer chain)
\cite{ivanov97,koga,schul02a}, that more general product eigenstates
containing chain fragments of finite length lead to an infinite series of
magnetization plateaus \cite{schul02}.

In this paper we discuss a recently discovered \cite{prl02} 
class of quite universal eigenstates of
frustrated quantum antiferromagnets which become groundstates in strong
magnetic fields and lead to macroscopic 
jumps in the magnetization curve close to
saturation. In what follows we consider in 
more detail the sawtooth chain and the checkerboard lattice.

\section{Localized magnon states}
\label{sec2}
We consider $N$ quantum spins of ``length'' $s$ described by the
Heisenberg Hamiltonian
\be
\label{ham1}
 \hat{H} = \sum_{ij } J_{ij}
\left\{{s}_i^x {s}_j^x + {s}_i^y {s}_j^y + 
{s}_i^z {s}_j^z  \right\}
       - h {S}^z.
\ee
${S}^z = \sum_i{s}^z_i$ is the $z$-component of the total spin,
$h$ is the magnetic field, and the
$J_{ij}$ are the exchange constants.

If the magnetic field $h$ is sufficiently strong $h \ge h_s$, 
the groundstate of
(\ref{ham1}) becomes the fully polarized ferromagnetic state 
(magnon vacuum state) $\vert 0 \rangle
=\vert\uparrow\uparrow\uparrow\ldots\rangle $
where
all spins assume their maximal $s_i^z$-quantum number and $M=\langle
S_z\rangle =Ns$.
The lowest excitations for $h > h_s$ are one-magnon
states $|1\rangle$ with $M=Ns-1$. They  
can be written as
$
    {}\vert 1\rangle \sim \frac{1}{c}\sum_{i}^{N}a_i {s}_i^-\vert 0\rangle,
$
where in general the excitation is distributed over the whole system.
However, for highly frustrated magnets having special bond geometry it 
turns out that the excitation can be localized over a restricted area $L$
of the
system, i.e. we have  
\be {}|1\rangle \; \to \;  |1\rangle_L 
 =\frac{1}{c} \sum_{i \in L} a_is_i^-|0\rangle
 =|\Psi_L\rangle|\Psi_R\rangle,  \label{eq2} 
\ee
where 
$|\Psi_L\rangle$ is the wave function of the magnon localized on area $L$
and 
$|\Psi_R\rangle$ the wave function 
of the fully polarized ferromagnetic remainder $R$ 
containing all
sites not belonging to $L$. In (\ref{eq2}) the constant $c$ is chosen to
normalize $|1\rangle_L$.
To demonstrate this
we split the Hamiltonian into three parts $\hat H= \hat H_L +\hat H_{L-R}
+\hat H_R$, where $\hat H_L$ ($\hat H_R$) contains only spins 
belonging to the local area $L$ (remainder $R$)
and $\hat H_{L-R}$ is the interaction term  between $L$ and $R$.
We restrict our consideration to the case that 
the wave function $|\Psi_L\rangle$ is an  eigenstate of 
$\hat H_L$ and has lowest one-magnon energy.
Of course,
the
wave function $|\Psi_R\rangle $ is an eigenstate of $\hat H_R$. 
Now we demand that the total wave
function (\ref{eq2}) is an eigenstate of the full Hamiltonian $\hat H_L+ 
\hat H_{L-R}+ \hat H_R$. 
After some manipulation  one finds that the exchange couplings 
in the interaction term 
$\hat H_{L-R}$ have to fulfill two conditions\cite{comment}, namely 
\be \label{eq3}
\sum_{l \in L} J_{rl} a_l = 0 \quad \forall \; r \in R  
\ee
and \be \label{eq4}
\sum_{r \in R} J_{rl} = const. \quad \forall \; l \in L.  
\ee
The first one comes from the xx and yy terms in $\hat H_{L-R}$ and the
second one from the zz term (i.e. condition (\ref{eq4}) is not relevant for
the pure xy model). 
Eq. (\ref{eq3}) leads to condition on the bond geometry, whereas
eq. (\ref{eq4}) is a condition for the bond strengths and 
is automatically 
fulfilled in uniform lattices with equivalent sites.
Indeed one finds that the above 
conditions are fulfilled for remarkably many different lattices/models
such as
the kagom\'e and the sawtooth chain, 
the 2d kagom\'e, square-kagom\'e or checkerboard lattices, 
the 3d pyrochlore lattice
but also for a fractal lattice like the Sierpinski gasket. 

In Fig.1 we illustrate the localized magnon states on the 
sawtooth chain with $J_2=2J_1$ and on the checkerboard 
lattices. Both 
systems attract 
currently a lot of attention as examples for novel low-energy physics in
quantum systems (see e.g. \cite{checker,sawtooth}). 
We show in the figures only the localized magnons of minimum size. On the 
checkerboard  lattice  localized magnons can sit on the four sites 
of an 'empty' square but also 
on the $\sqrt{N}$ sites of a 
vertical, horizontal or sloping (45$^\circ$) line. On the sawtooth
chain a magnon can sit on the 
three neighbouring sites forming  '{\bf V}' but also 
on the $N/2$ sites on the base line.

\begin{figure}
\begin{center}\epsfxsize=26pc
\epsfbox{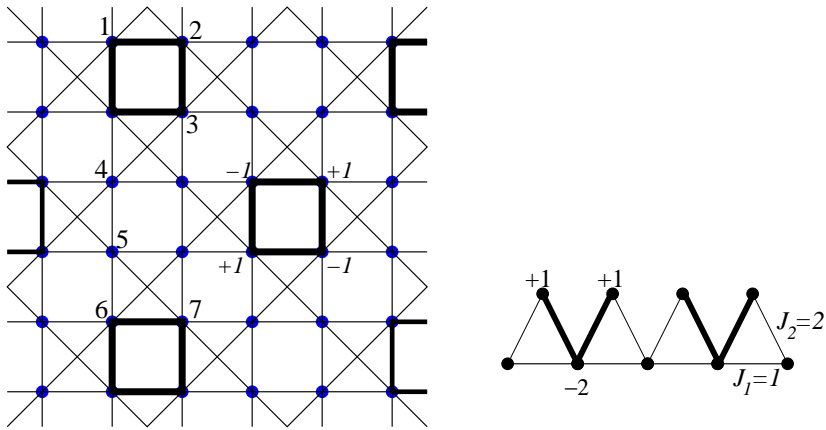}
\end{center}
\caption{\label{fig1} Localized magnon states on the checkerboard lattice
 and on the sawtooth chain with $J_2=2J_1$. 
The position of the localized magnons 
are indicated  by extra thick
lines. The numbers +1, -1, -2 at the corners represent the coefficients $a_i$ 
(see eq. (\ref{eq2})). }
\end{figure}
The next step is to construct eigenstates containing $n > 1$  localized
magnons on different localized areas $L_\alpha$, $\alpha=1,\ldots,n$. 
If these localized areas are separated from each other one can still fulfill
both conditions
(\ref{eq3}) and (\ref{eq4}). This way one can find a whole class of product
eigenstates of the form $|\Psi \rangle = |\Psi_{L_1}\rangle
|\Psi_{L_2}\rangle \cdots 
|\Psi_{L_m}\rangle  |\Psi_R\rangle$, where the 
$|\Psi_{L_\alpha}\rangle$
and $|\Psi_{R}\rangle$ are defined in analogy to  (\ref{eq2}). 
The system can
be filled with separated magnons as long as  the magnons do not interact.
The maximum number of localized magnons depends on 
the lattice geometry. The state with maximum filling corresponds to a 
'magnon
crystal', i.e. to a state with a regular arrangement of magnons (see
Fig.\ref{fig1}). The 'crystalline' magnon state of the sawtooth chain
(checkerboard lattice)  is twofold (fourfold)
degenerated and  breaks spontaneously the translational symmetry of the lattice.
Maximum filling for the checkerboard lattice 
is reached if magnons sit on 
every forth 'empty' square and for the sawtooth chain 
if magnons sit on 
every second '{\bf V}'. Each localized magnon lowers the magnetic 
quantum number $M$
of total $S_z$ by one, i.e. 
one has  $M=Ns-n$ for a state with $n$ localized magnons.
Hence  the corresponding quantum number of the 
'magnon crystal' is $M=Ns-N/8$
(checkerboard) and $M=Ns-N/4$ (sawtooth).

Due to the simple product form of the eigenstates one can calculate
explicitely the spin-spin correlation functions. They depend on the
parameters $a_i$. One has three different types of correlations, namly
within $R$, within $L$ and between $L$ nad $R$.  
As an example we give the correlations 
for the checkerboard lattice with localized magnons on
'empty' squares: 
$\langle \vec{s}_1\vec{s}_2\rangle = - 1/4$; $\langle {s}^z_1{s}^z_2\rangle =
0$;  $\langle \vec{s}_1\vec{s}_3\rangle =  1/4$; $\langle {s}^z_1{s}^z_3\rangle =
0$; $\langle \vec{s}_1\vec{s}_4\rangle = \langle {s}^z_1{s}^z_4\rangle =
1/8$;   $\langle \vec{s}_1\vec{s}_6\rangle = \langle {s}^z_1{s}^z_6\rangle =
 \langle \vec{s}_1\vec{s}_7\rangle = \langle {s}^z_1{s}^z_7\rangle = 1/16$
and $ \langle \vec{s}_4\vec{s}_5\rangle = \langle {s}^z_4{s}^z_5\rangle =
1/4$ (the numbers correspond to those given in Fig. \ref{fig1}). Other
correlation functions can be obtained  by symmetry arguments having in mind
that there is no real distance dependence of correlations within $R$ and
between $L$ and $R$.

Finally,  we mention that the existence of localized magnon states in
regular spin lattices is related to the existence of flat bands in the
magnon dispersion\cite{prl02}.    

\section{Macroscopic magnetization jump}
The interest in the above described eigenstates is not only an academic one.
One can show rigorously that these eigenstates under certain conditions 
become groundstates in a
magnetic field \cite{schmidt}. This is true for instance in many
systems with translational symmetry.
We define as magnetization 
$
m = \frac{\left \langle S^z \right \rangle }{Ns} = \frac{M}{M_0}
$,
i.e. $m$ is normalized to unity for the fully polarized ferromagnetic
state $|0\rangle$.
\begin{figure}
\begin{center}\epsfxsize=17.7pc  \epsfbox{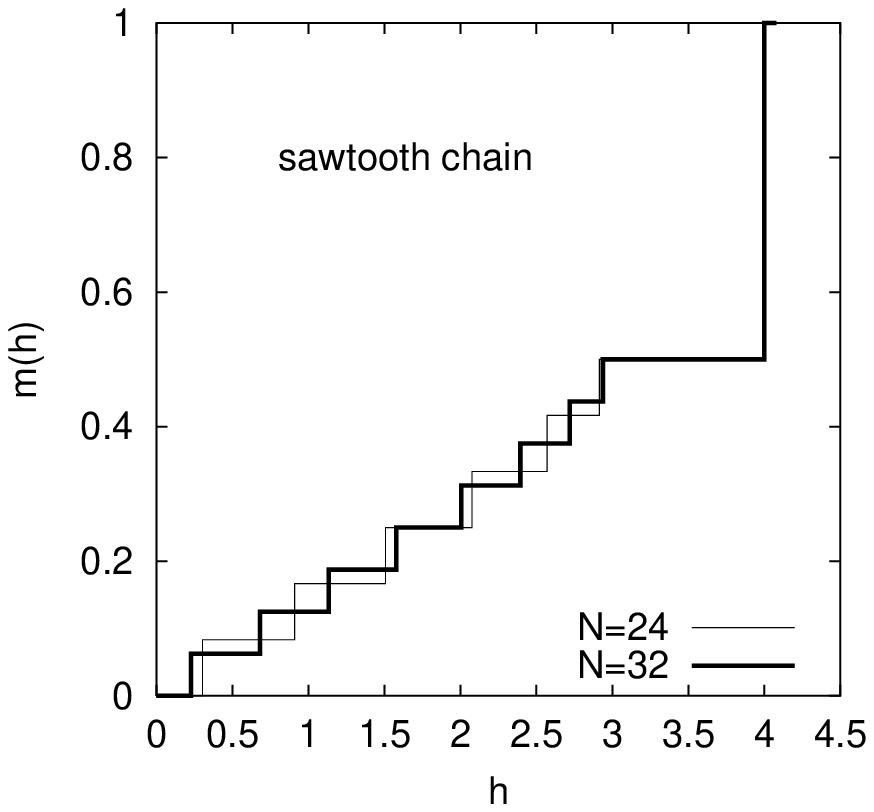}
\epsfxsize=17.7pc \epsfbox{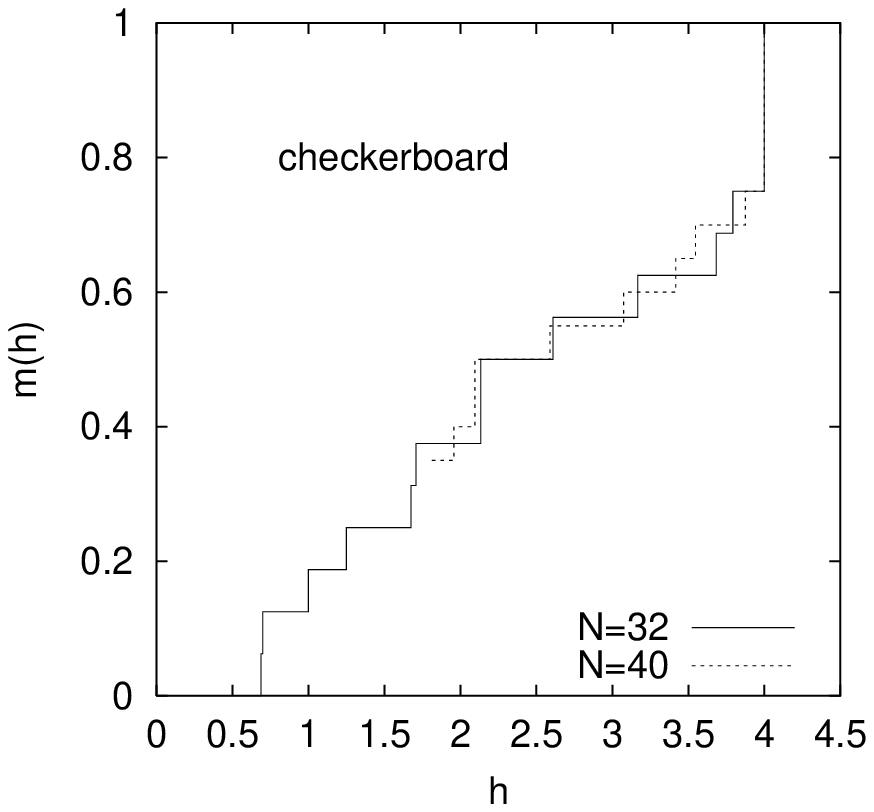}
\end{center}
\caption{\label{fig2}Magnetization $m$ versus magnetic field $h$ of the
spin half Heisenberg antiferromagnet  on the
sawtooth chain ($J_2=2J_1$, cf. Fig. \ref{fig1}) and on the 
checkerboard lattice. For the finite
checkerboard lattice of $N=40$ only the upper part of the curve was
calculated.
}
\end{figure}
Since the z component of the total spin commutes with the Hamiltonian
one can calculate the magnetization
from the lowest zero-field 
energies  $E(M)$ in each sector of $M$, i.e.
$E(M,h) = E(M) - hM $ and  $h(M)=\frac{\Delta E}{\Delta M}$. 
The eigenstates discussed in section \ref{sec2} contain localized
noninteracting magnons and have therefore well defined energies 
$E(M=Ns-n)= Ns^2-n\varepsilon_1$, where $\varepsilon_1$ is 
the energy difference between the
magnon vacuum  and state with {\bf one} magnon and $n$ is the number of
localized magnons. Hence we have a straight part in the $E$ versus $M$ curve
close to saturation leading to a jump in the $m(h)$ curve.
The height of the jump $\Delta m$ is determined by the maximal number 
$n_{max}$ of 
localized magnons in the system and on the  spin 'length' $s$,
$\Delta m =  n_{max}/Ns$. Obviously the jump is a true quantum effect and
vanishes for $s \to \infty$.

In Fig. \ref{fig2} we show magnetization curves for the sawtooth
chain and the checkerboard 
lattice for $s=1/2$ calculated with exact diagonalization for finite 
lattices. The height of the jump for the sawtooth chain is 
$\Delta m=1/2$ and for the checkerboard lattice  
$\Delta m = 1/4$. We emphasize
that $\Delta m$ does not depend on the size of the system provided the
periodic boundary conditions of the finite lattice fit to the translational
symmetry of the 'magnon crystal'. 
For the sawtooth chain we see a well-pronounced plateau preceding the jump.
This plateau belongs to the 'magnon crystal' state.
Though such a plateau is less pronounced for the checkerboard lattice there are
general arguments\cite{plat} that such a crystalline magnon state should
have gapped
excitations and may  therefore be connected with a magnetization plateau at
$m=1-\delta m = 1 - n_{max}/Ns$.    
We mention that the plateaus fulfill the condition 
of 
Oshikawa, Yamanaka and Affleck
\cite{oshikawa} derived
for plateaus in 1d
systems  not only for the sawtooth chain but also for the 2d checkerboard
lattice.

To our knowlegde the jump was not observed experimentally so far. 
To see 
the jump and the preceding plateau in experiments one needs
highly frustrated magnets with small spin quantum number $s$ and sufficiently
small exchange coupling strength $J$ to reach  the saturation field.
\\

\ack

This work was  supported by 
the Deutsche Forschungsgemeinschaft (Grant No. Ri615/10-1).


\section*{References}
\begin{thebibliography}{99}
\bibitem{mattis} Mattis D C {\it The Theory of Magnetism I}, Springer, Berlin
1988
\bibitem{bethe} Bethe H A 1931  {\it Z.\ Phys.} {\bf 71} 205
\bibitem{majumdar}  Majumdar C K and  Ghosh D K 1969  \JMP 
\ {\bf  10 } 1399 
\bibitem{shastry} Shastry B S and Sutherland B 1981  {\it Physica} 
{\bf 108B} 1069 
\bibitem{cugeo} Poilblanc D, Riera J, Hayward C A, Berthier C and
Horvatic M 1997
\PR B {\bf 55} 941
\bibitem{srcubo} 
    Miyahara S and Ueda K 1999  \PRL {\bf 82},
    3701 
\bibitem{pimpinelli} Pimpinelli A 1991 
 \JPCM {\bf 3} 445 
\bibitem{ivanov97} Ivanov N B and Richter J 1997 
      \PL A {\bf 232} 308 ;
      Richter J,  Ivanov N B and Schulenburg J 1998
      \JPCM {\bf 10} 3635 
\bibitem{japaner3d} 
Ueda K and Miyahara S 1999 \JPCM  {\bf 11} L175 
\bibitem{hamada} Hamada T, Kane J, Nagawaka S and Natsume Y \JPSJ 1988
{\bf 57} 1891      
\bibitem{koga} Koga A, Okunishi K, and Kawakami N 2000
 \PR B {\bf 62} 5558 ; Koga A and Kawakami N 2002 \PR B
  {\bf 65}, 214415
\bibitem{schul02a} Schulenburg J and Richter J 2002 
 \PR B {\bf 66} 134419
\bibitem{schul02} Schulenburg J and Richter J 2002
 \PR B {\bf 65} 054420
\bibitem{prl02} Schulenburg J, Honecker A, Schnack J, Richter J  
and Schmidt H-J 2002 \PRL {\bf 88} 167207 
\bibitem{comment} The second condition is not a necessary one, i.e. one can
find models  with eigenstates of form (\ref{eq2}) violating (\ref{eq4}), see 
\cite{prl02}.  
This more general case appears if 
$| \Psi_L\rangle| \Psi_R\rangle$ is not an individual 
eigenstate of both $\hat H_L$
{\bf and} $\hat H_{L-R}$ but of $(\hat H_{L-R}+\hat H_{L})$. 
\bibitem{checker}Palmer S E and Chalker J T 2001 \PR B
{\bf 64}, 094412 ;  Brenig W and Honecker A 2002 \PR B {\bf 65} R140407 ; 
Fouet J B, Mambrini M, Sindzingre P and Lhuillier C
 2003 \PR B {\bf 67} 054411 
\bibitem{sawtooth}
 Sen D, Shastry B S, Walstedt R E, Cava R 1996  \PR B 
{\bf 53} 6401 ;
Pati S K 2003 \PR B {\bf 67} 184411 ;
Chandra V R, Sen D,   Ivanov N B  and Richter J 2003 {\it cond-mat/0307492}
\bibitem{schmidt}
Schnack J, Schmidt H-J, Richter J  and Schulenburg J 2002 {\it Eur.
Phys. J.} B 2001 {\bf 24} 475 ;  Schmidt H-J 2002 \JPA 
{\bf 35}  6545
\bibitem{plat}
Momoi T and Totsuka K 2000 \PR B {\bf 61} 3231 ;
 Oshikawa M 2000 \PRL {\bf 84} 1535
\bibitem{oshikawa}    
  Oshikawa M, Yamanaka M and Affleck I 1997 \PRL {\bf 78} 1984



\end{thebibliography}
\end{document}